\documentclass{article}
\usepackage{epsfig}
\usepackage{amssymb}
\usepackage{amsmath}
\usepackage{amsfonts}
\newcommand \be{\begin{eqnarray}}
\newcommand \ee{\end{eqnarray}}
\begin{document}
\begin{center}
{\bf Spin 1/2 Fermions in the Unitary Limit.III}\\
\bigskip
\bigskip
H. S. K\"ohler \footnote{e-mail: kohlers@u.arizona.edu} \\
{\em Physics Department, University of Arizona, Tucson, Arizona
85721,USA}\\
\end{center}
\date{\today}

\begin{abstract}
In scattering theory, the unitary limit  is defined by an infinite scattering
length and a zero effective range, corresponding to a phase-shift
$\pi/2$, independent of energy.
This condition is satisfied by a rank-1 separable potential $V(k,k')=-v(k)v(k')$
with $v^{2}(k)=
(4\pi)^{2}(\Lambda^{2}-k^{2})^{-\frac{1}{2}}$, $\Lambda$ being the cut-off
in momentum space.
Previous calculations using a Pauli-corrected ladder summation to calculate
the energy of a zero temperature 
many body system of spin $\frac{1}{2}$ fermions with this interaction
gave $\xi=0.24$
(units of kinetic energy) \it independent of density \rm 
and with $\Lambda\rightarrow\infty$. 
This value of $\xi$ is appreciably smaller than the experimental and that 
obtained from other calculations, most notably from Monte Carlo, which 
in principle would be the most reliable. Our previous work did however also show a
strong dependence on effective range $r_0$ (with $r_0=0$ at unitarity).
With an increase to $r_0=1.0$ the energy varied from  $\xi \sim 0.38$ 
at $k_f=0.6 fm^{-1}$ to $\sim 0.45$ at
$k_f=1.8 fm^{-1}$  which is somewhat closer to the Monte-Carlo results.
These previous calculations are here extended by including
the effect of the previously neglected mean-field propagation, the
dispersion correction.  This is repulsive and 
found to increase drastically with decreasing effective range.
It is large enough to suggest a revised value of $\xi \sim
0.4\Leftrightarrow 0.5$ independent of $r_0$. 
Off-shell effects are also investigated  by introducing
a rank-2 (phase-shift equivalent) separable potential. Effects 
of $10\%$  or more in energy could be demonstrated for $r_0>0$.
It is pointed out that a computational cut-off in momentum-space 
brings in another scale in the in principle scale-less unitary problem. 
\end{abstract}
 
\section {Introduction}
The energy of a spin-$\frac{1}{2}$ fermion gas  in the
unitary limit is of current experimental and theoretical interest as
evidenced by the many publications on this subject matter.

Our earlier results  were presented in two previous papers
\cite{hsk07,hsk08}.  The total energy in units
of the kinetic energy was found to be $\xi=0.24$ independent of density.
Other calculations as
well as experiments show appreciably higher values, $\xi \sim .45$.
\cite{geh03,bar04,car03,per04,hei01,bak99,siu08,don10}
This discrepancy is of considerable interest  from a many-body 
theoretical point of view and is the motivation for this paper.

There are two distinctly different sources  that determine the result of
any many-body calculation. The first is the interaction the second the
many-body theory used.
If assuming a separable interaction 
the inverse scattering formalism allows for the construction of 
interactions (in principle an infinity number) that fit any given set of
phasehifts exactly. If the phases do not change sign (and no coupling
between
states exist) a rank-1 potential is sufficient for a complete numerical
fit. 
In the unitary limit, defined by an infinite scattering
length, $a_s\rightarrow \infty$, and
effective range $r_0=0$, the rank-one separable interaction $<k|V|k'>=v(k)v(k')$
fitted to phase-shifts $\delta(k)=\frac{\pi}{2}$ for $k\leq\Lambda$
is given by \cite{hsk07}:

\begin{equation}
v^{2}(k)= \frac{(4\pi)^{2}}{(\Lambda^{2}-k^{2})^{\frac{1}{2}}}
\label{vpi2}
\end{equation}

In the limit $\Lambda\rightarrow\infty$ this interaction
reduces to a delta-function in
coordinate space. Any numerical calculation necessitates a cut-off in
momentum-space and as shown by eq. (\ref{vpi2}), this requires the strength
to be increased with increasing $k$ and a singularity at $k=\Lambda$,
that in general makes this interaction difficult to handle numerically. 
An analytic result for the total energy could however be obtained in a
specified approximation \cite{hsk07} yielding the value quoted above,
$\xi=0.24$, independent of density. 
A strong dependence on the effective
range was however also reported. It was  for example found that
if increasing $r_0\rightarrow 1fm$,   $\xi$ increases to $\sim 0.4$ 
close to other reports  (see citations above). 

\begin{figure}
\centerline{
\psfig{figure=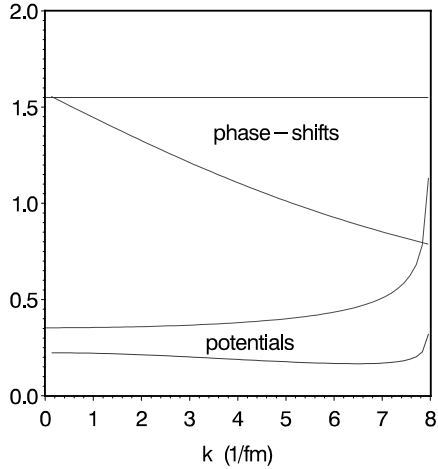,width=7cm,angle=0}
}
\vspace{.0in}
\caption{There are two sets of curves. The upper are phase-shifts the lower
the corresponding rank-1 potentials defined by $v(k)$. 
The uppermost in each set is for the
unitary case, $a_s\rightarrow\infty$ and $r_0=0$, the lower in each set is
for the 'near' unitary case with $r_0=0.25$.
Unit is Fermis ('fm').
}
\label{wavnewpot}
\end{figure}

Calculations, other than the 'exact' at $r_0=0$ were found to be
computationally difficult for $r_0<0.25$. This situation can be understood
by studying Fig. \ref{wavnewpot},
showing a big difference in phase-shifts between the two cases $r_0=0$
and $0.25fm$ respectively. This figure also shows that the corresponding 
potentials are dramatically different
because of the singularity at $k=\Lambda=8 fm$ in the unitary limit.

Any fit to the
on-shell phase-shifts alone does of course not define the interaction
uniquely except exactly at the unitary limit as argued below.
But one advaantage of the inverse scattering method is that off-shell
properties can easily be monitored and changed  
by increasing the rank of the potential while maintaining 
the on-shell fit to phase-shifts.
Results of such a modification of the interaction will be shown below.

The most accurate many-body theoretical method is at least in principle
the Monte Carlo, although it may be hampered by computer-technical limitations.
The Brueckner nuclear many-body theory was the first to (with considerable
success) describe nuclear matter saturation and finite nucleus properties.
This theory defines an effective in-medium interaction which 
can be regarded as a modification of the  free scattering
$T$-matrix:
\begin{equation}
T=v+v\frac{1}{e_0+i\epsilon}T
\label{T}
\end{equation} 
or more relevant
the reactance matrix $\cal K$:
\begin{equation}
{\cal K}=v+v\frac{\cal P}{e_0}{\cal K}
\label{K}
\end{equation} 
where ${\cal P}$ indicates the principal value and $e_0$ kinetic energies.
These free scattering interactions are modified by the definition of an
in-medium effective interaction, the Brueckner
Reaction Matrix involving  two important modifications that can be
physically justified. The 
first recognizes the statistics of a Fermion gas and the resulting
Pauli-blocking in the many-body system. The second modification was 
in accordance with a, at the
time of incipience of this theory,  new idea, the mean field concept,
a corner stone in the theory of both the
shell and the optical model. These two effects are included
self-consistently in the Brueckner Reaction Matrix $K$ defined by:
 
\begin{equation}
K=v+v\frac{Q}{e}K
\label{BK}
\end{equation}
where $e=e_0+ U$.
The particle propagator now includes interactions with neighboring
nucleons via the mean field $U$ defined self-consistently by the in-medium
$K$-interaction. As further detailed in Sect. (2) this  is 
in effect a 3-body term. 
The Brueckner theory  as defined by the effective interaction (\ref{BK}) 
implies a summation of
a particular sub-set
of perturbation terms (diagrams).
Considerable effort has in the past been
made to estimate corrections to this but they were all relatively small
(or too uncertain to be regarded) and rarely included in what is
commonly referred to as "Brueckner Theory".
This theory  was primarily designed to be applied to nuclei.
There is no \`a priori reason why it should be
applicable to the problem at hand with a substantially different
interaction.
(Even though  neutron matter
sometimes is quoted as being 'close' to a unitary gas, our results suggest it to
be a poor approximation.)
Although the estimates of higher order terms in nuclear matter calculations
show them to be relatively small, these terms may be important in the
unitary system because of  the rather different interaction.
This is however a question beyond the scope of this investigation. 

From a many-body theoretical point of view it is rather of interest to find
if Brueckner theory as formulated above
will agree with Monte Carlo and other reported results for the energy of a
unitary gas.

Our previous calculations with the rank-1 potential at or near the unitary
limit referred to above were made using Brueckner
theory, but neglecting the mean-field $U$ above,  i.e. with $e\equiv e_0$.

The effect of including the mean field in hole- (but not particle-)
line propagation  is now also 
investigated with results shown below in Sect. 2.
Sect. 3 deals with effects of off-shell modifications, while Sect. 4
contains a summary and some conclusions.

\section{Mean-field effects}
In the many-body theory of nuclei the effect of nucleons propagating
not as free but rather as bound in a shell-model potential (the mean field)
is extremely important and is
a major contributor to nuclear saturation and finite nucleus stability
against collapse. This effect is included in the Brueckner $K$-matrix
by   using $e=e_0+U$ rather than just $e_0$ (kinetic energy) in eq. 
(\ref{BK})
and is in the literature referred to as a \it dispersion \rm correction.
It can be estimated as follows. 
Let $\Delta U$ be the averaged difference
between hole and particle potential energies. The dispersion correction
$\Delta K_{disp}$ is
then found by differentiating $K$ in eq. (\ref{BK})
by the energy denominator $e$ (or $U$) to get \cite{mos60,hskm07} 

\begin{equation}
\Delta K_{disp} \propto \Delta U*I_{w}.
\label{disp}
\end{equation}

where the wound-integral $I_{w}$ is defined  by
$$I_{w}=\int (\Psi(r)- \Phi(r))^{2}d{\bf r}$$
with $\Psi$ and $\Phi$  the {\bf correlated} and {\bf uncorrelated}
two-body  wave-functions respectively.
It is relatively small at low density (small
finite nuclei) but grows with density because of the increased binding.
And it is repulsive which explains why it is an important
contribution to saturation. It is basically a three-body effect
as the effective two-body interaction depends on
the presence of "third nucleons" that constitute the mean field $U$. And
it is important that in nuclei, $U$ is momentum-dependent (non-local) so
that $\Delta U \neq 0$.
The wound-integral $I_{w}$ is an important quantity by itself being a
measure of the correlation strength.

Calculations have shown that in nuclei this dispersion 
effect is less important for
the $^1S_0$ interactions but more so for the $^3S_1-^3D_1$
interactions. Eq (\ref{disp}) shows it to be
proportional to both $\Delta U$ and $I_w$. 
The $^1S_0$ interaction is defined (partially) by a
scattering length  and an effective range  $a_s=-18.5 fm$ and $r_0=2.8 fm$.
The reason the dispersion effect is
small for this state in nuclei is that $I_w$ is small for this interaction. 
It is appreciably
larger in the coupled $^3S_1-^3D_1$ state. It is found below that 
approaching the unitary limit
with $a_s \rightarrow \infty$ and $r_0 \rightarrow 0$, $I_w$ will increase.
But $\Delta U$ will decrease because the mean-field becomes less
momentum-dependent so that the difference between the hole and the 
particle potential energies becomes small.
The latter effect  motivated the neglect of the dispersion effect
in the Brueckner-equation in our previous unitary calculations. 

In the results presented in this report the mean
field \it is \rm included, but \it only for hole-propagation. \rm
The justification for this choice goes as follows. 
The difference between mean-field interactions of 
nucleons propagating as holes or as particles
respectively was shown already by Brueckner and Gammel \cite{br57,hsk75}
and further investigated in many
papers e.g. \cite{hsk73}. In most later nuclear calculations one has however, 
without strong mathematical proof, reverted
to treating the hole and particle propagations on an equal footing. (The 
"continuous" choice).
As emphasized by Bethe\cite{bet65,bet71} these insertions are basically 
3-body interactions and can be treated as such by the Faddeev method as he
also did. 

The mean field in the nuclear medium has a definite momentum-dependence with
$m^*\sim 0.8$. This is not the case in a gas in the unitary limit. 
Although important in nuclear many body theory 
there is therefore no  \`a priori reason why the inclusion of  the hole-line
interactions alone should provide a better approximation for the energy of the
unitary gas. It seems however worth-while to investigate this  effect 
in more detail.

\begin{figure}
\centerline{
\psfig{figure=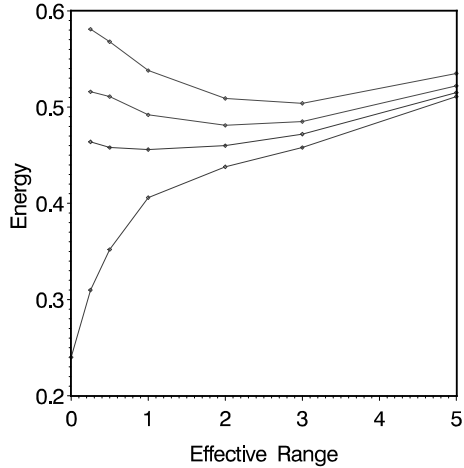,width=7cm,angle=0}
}
\vspace{.0in}
\caption{Energy, in units of kinetic energy, is shown as a function of effective
range $r_0$ (in units of $fm$). Scattering length is $a_s\rightarrow\infty$. 
The lowest curve shows the energy
calculated without any insertion in hole lines. The remaining three curves 
show, from top to bottom, the results with insertions being $1.0,\frac{1}{2}$, and
$\frac{1}{4}$  $\times U_h(k)$ respectively, where $U_h$ is the selfconsistently
calculated mean field for hole lines, i.e. for $k \leq k_f$.
The fermimomentum is $k_f=1 fm^{-1}$ as in all calculations in this work. 
}
\label{unir0}
\end{figure}

Some results of our calculations are shown in Fig \ref{unir0}.
The  lowest curve  is for free propagation without a mean field 
($U\equiv 0$) at $k_f=1 fm^{-1}$. (Fig. 3 in ref. \cite{hsk08} shows 
results also at other densities with $U\equiv 0$.) 
Note that the point at $r_0=0$, the unitary limit, taken from the previous
(mostly) analytic calculation in ref. \cite{hsk07}
is reached by smooth extrapolation from the data obtained in 
the present calculations.
The upper three curves show that the effect
of hole-line insertions, the dispersion correction, as mentioned above is
repulsive.
One furthermore sees that this repulsion \it increases \rm with \it decreasing \rm
effective range. In accordance with eq. (\ref{disp}) this is associated
with an increase in the wound-integral $I_w$, i.e. correlations, with
decreasing effective range.  The mean field $U_h$ is on the other hand
found to be essentially independent of $r_0$.
The important message of the results shown here is the rather dramatic 
change in the $r_0$ dependence of $\xi$ nearly nullifying it for
$U(k)=\frac{1}{4}U_h(k)$.
Further comments on these results are found in Sect. 4.

\begin{figure}
\centerline{
\psfig{figure=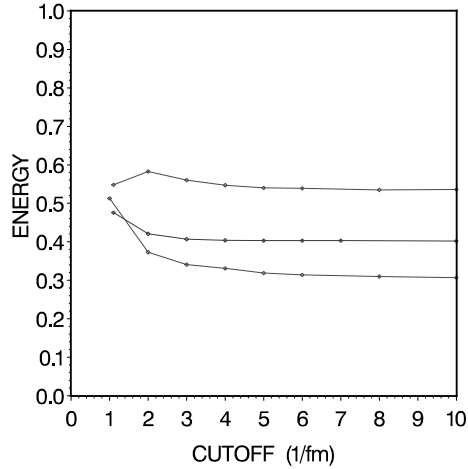,width=7cm,angle=0}
}
\vspace{.0in}
\caption{The middle curve shows the energy, in units of kinetic energy without
insertions in hole lines, i.e. only kinetic energy in the propagators.
The
upper curve includes the self-consistent mean field $U_h$ (see text). 
The horizontal axis is labelled with the cutoff $\Lambda$ in units of
$fm^{-1}$. The interaction is for these two curves defined by $r_0=1 fm^{-1}$ and
$a_s\rightarrow \infty$ and a rank-1 potential.
The lowest curve is similar to the middle except that the effective range
now is $r_0=0.25 fm$, i.e. somewhat closer to the unitary limit for which $r_0=0$.  
}
\label{unicut}
\end{figure}

The results shown in Fig. \ref{unir0} (as in most results reported here),
were made with a  cut-off  in momentum space  $\Lambda=10 fm^{-1}$.
Fig. \ref{unicut}, on the other hand shows the near independence of $\Lambda$.
 The middle curve, in particular, shows this to be so
for $\Lambda>2k_f$, i.e.  the 'range' of the
$Q$- (Pauli-)operator. There are no insertions ($U\equiv 0$)
in the propagator lines in this curve.
The upper curve does include the repulsive effect of hole-line insertions. 
It has a slight bump at smaller cut-offs.
A very similar result was shown in a previous work (\cite{hskm07}) for nuclear
forces.
The unitary limit is unique. The only energy-scale is the kinetic energy.
This was explicitly shown to be the case in the ladder-approximation with
the interaction of eq. (\ref{vpi2}) where the 
quantity $\xi$ was shown to be
independent of density as is expected in the unitary limit. \cite{hsk07}
In that calculation the limit $\Lambda\rightarrow\infty$
could be reached explicitly and exactly which made this calculation unique.
Fig. 3 in ref.
\cite{hsk08} on the other hand,  shows a definite density-dependence 
when $r_0\neq 0$, even though $a_s\rightarrow \infty$.

A finite value of $\Lambda$ does however bring in another scale in theses
calculations.
It is evident that for any fixed value of $\Lambda$ a density-dependence 
has to exist, because for 
$k_f>\Lambda/2$ (i.e. $\Lambda<2k_f$) the solution of the reaction-matrix
$K$ would  involve momenta that are larger than those restricted by
$\Lambda$. This explains the rapid increase in 
energy for $\Lambda \leq 2-3 fm^{-1}$ shown in
Fig. \ref{unicut}. Only for appreciably larger values of $\Lambda$ does one
see a constancy at least in case of $r_0=1 fm$. The
smallest value of $r_0$ for which reliable calculations could be made here
was $r_0=0.25 fm$. The $\Lambda$-dependence for this value of $r_0$ is shown
by the lowest curve in Fig. \ref{unicut}. It does indeed show a larger but
maybe not significant increase in $\Lambda$-dependence.

\section{Off-shell scattering}
A potential interaction fitted to scattering phase-shifts by
inverse scattering or any other method, is not unique. This 
is of course a reason why so much effort has
been expanded to construct meson-theoretical and QCD derived
NN-interactions for use in many-body calculations.
The phase-shifts are on-shell data. The inverse scattering method is based on
inverting the reactance-matrix $\cal K$ of eq. (\ref{K}) with the diagonal
elements given by $<k|{\cal K}|k>=\tan \delta(k)$. 
In any theory that has the purpose of explaining the properties of a
many-body system from two-body data
this on-shell diagonal information would be  necessary but in general
not sufficient. 
In any many-body theory and specifically in the case at hand, Brueckner
theory, off-diagonal elements are needed to
construct the in-medium off-shell interaction, the reaction matrix
defined by eq.(\ref{BK}).
No direct experimental off-shell
data are however in general available for this purpose as it  is not an
observable\cite{fur01} and in fact indistinguishable from many body forces. 
Their
relative contributions ('strengths') are indistiguishable and not subject
to observation referred to as 'the equivalence theorem'.

Having said that, the problem still remains: To what extent, numerically
does the result of a many-body calculation depend on variations in
off-shell propagation.
Such a study is very conveniently done within the inverse
scattering formalism. To demonstrate this, consider the phase-shifts
defined by 
some scattering length and effective range. If they
are all of the same sign  they can (easily) be reproduced by a rank-1
separable potential.
The low-energy ($E<150 MeV$) nuclear $^1S_0$ interaction 
is a good example and a rank-1 potential is in this case a "good" 
representation. 
 This is exemplified by the close numerical agreement
with the Bonn off-diagonal elements shown in Fig. 3 of ref. \cite{kwo95}. It
can be understood to be a consequence of
the appearance of a pole in the scattering matrix near the real axis which
is the case for potentials with large scattering legth and small effective
range. 
For other states, in particular the
$^3P_1$ a rank-1 potential is sufficient to fit the phase-shifts but
it disagrees with results of the Bonn-potentials and this can be traced to
a difference in off-diagonal reactance matrix elements. \cite{kwo95}

Off-diagonal elements can in fact easily
be changed, while preserving the on-shell phase-shift information, by 
increasing the rank. It was shown already by Chadan\cite{cha58} and by
Fiedeldey\cite{fie69}  how to do this. 
The method assumes a given set of phase-shifts $\delta(k)$
and an  arbitrarily chosen potential with some associated
phase-shifts $\delta_0(k)$. A
second potential can then be calculated so that the sum, a rank-2 potential
reproduces the given set $\delta(k)$. This method was used by Chadan to
show the effect of varying off-shell behavior. (The method was extended by
Fuda\cite{fud70} to coupled channels and used in a previous work by Kwong and the
author \cite{kwo95}).

\begin{figure}
\centerline{
\psfig{figure=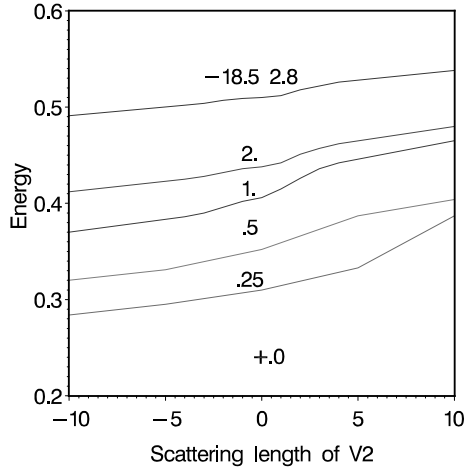,width=7cm,angle=0}
}
\vspace{.0in}
\caption{The energy, in units of the kinetic energy, is shown as a function
of the scattering length $a_s$ of potential $V_2$ of a rank-2 potential as
described in the text. 
The phase-shifts $\delta_0(k)$ are defined by this value of $a_s$ while
$r_0=\pm 10$ with the sign opposite to that of $a_s$.. 
The 'given set of phase-shifts', $\delta(k)$ is in case of the top curve
defined by
$a_s$ and $r_0$ as indicated, while for the lower curves only $r_0$ is
indicated while $a_s\rightarrow \infty$. The cross indicates the previously
reported result in the unitary limit $r_0=0$.\cite{hsk08}
Unit is Fermis ($fm$).
}
\label{unioff}
\end{figure}

This method to study the effect of off-shell variation is also used here and 
results are shown in Fig. \ref{unioff}. In the uppermost curve the given
set of phase-shifts
was defined by the nuclear $^1S_0$ scattering length $a_s=-18.5 fm$
and effective range $r_0=2.8 fm$, while the four lower curves show results
for 
$a_s\rightarrow\infty$ with $r_0=2,1,.5$ and $.25 fm$ as indicated in the diagram. 
The  second part
of the rank-2 potential $V_2$, 'arbitrarly chosen', was 
defined by $r_0=\pm 10$ while $a_s$ is given
by the numbers along the horizontal axis, i.e a potential of relatively 
short range in momentum-space. 
We wish to emphasize that every point along each of these curve 
are equivalent in the sense that the
potential at each point
all have the same scattering same phase-shift. They are phase-shift
equivalent. 
Note that $V_2\equiv 0$ when the scattering length $a_s=0$, and the slope is a
measure of an 'off-shell dependence'. Regarding for example the $r_0=0.25$
curve ($a_s\rightarrow\infty$) one finds $0.28<\xi<0.38$; there is no unique
answer to what the  energy is here, if phase-shifts are the only
information used to define the interaction.

The diagram illustrates the point that the energy (and other
properties) of a many-body system is not defined by on-shell propeties
alone; the potential is not defined solely by the phase-shifts. 
It is well-known  that the off-shell effect is 
equivalent to (indistinguishable from) that of many-body forces so that the
curves could also be interpreted as being functions of the strength of some
(unspecified) 3-body force.
It is however important to realize that there is no exactly solvable N-body 
theory for N>3. Higher order terms could also change with off-shell. 
There might even be cancellations.

In contrast, the three-body problem
e.g. the triton, is exactly solvable and off-shell (three-body) effects can
be calculated  (see for example ref. \cite{hsk09}.)

The results shown in Fig. \ref{unioff} are  for $r_0 \geq 0.25$. For
smaller values the calculations become inaccurate related to the
singular nature of the interaction in the unitary limit as shown in
eq. (\ref{vpi2}) and illustrated in Fig. \ref{wavnewpot}.
In this limit the main part of the calculation can however be done
analytically if $V_2=0$ and the result is $\xi=0.24$ as reported in ref\cite{hsk08}
and shown by the cross in Fig. \ref{unioff}.

The results shown in Fig. \ref{unioff} show a definite off-shell dependence
even for the smallest value of $r_0$. In the unitary limit $r_0\rightarrow
0$ such a dependence should not exist in an exact calculation but may of
course still be seen in the ladder-approximation used here. 
unitary interactions used in at least most calculations are only
approximate in that the limit $r_0=0$ is not satisfied. One purpsoe
of the present work is to exemplify that this may lead to uncertainties
e.g. in the determination of $\xi$
also due to  unknown and unspecified off-shell properties. 

\begin{figure}
\centerline{
\psfig{figure=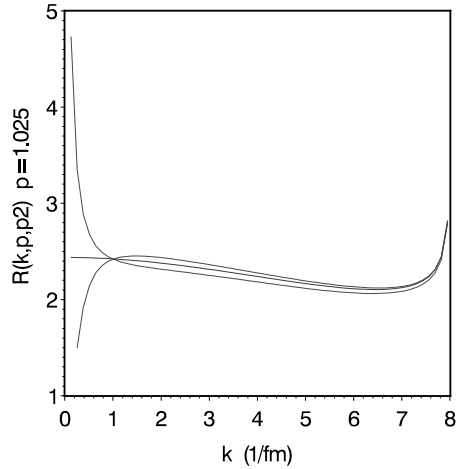,width=7cm,angle=0}
}
\vspace{.0in}
\caption{Half off-shell elements of the reactance matrix (eq. (\ref{K})
for three phase-shift equivalent potentials as detailed in the text.
Unit is Fermis ('fm').
}
\label{wavnewoff}
\end{figure}

For further illustration Fig. \ref{wavnewoff} shows half-shell 
reactance matrix elements.(See eq. (\ref{K})).
The middle curve is obtained with a rank-1 potential $V_1$ 
fitted to phase-shifts defined
by a scattering length $a_s\rightarrow \infty$ and an effective range
$r_0=0.25$ while $V_2\equiv 0$, i.e. the same parameters as for the lowest
curve in Fig. \ref{unioff} at the abscissa point =0.
The upper and lower curves are with rank-2 potentials with 
the same parameters as at the endpoints of that same curve in Fig.
\ref{unioff}.
The three curves differ in particular at small momenta, a consequence of
the short range (in momentum-space) of $V_2$.
The three curves intersect at the diagonal point $k=p=1.025 fm^{-1}$
because the three different potentials all fit the same
phase-shifts and the diagonal of the reactance matrix is ${\cal
K}=\tan(\delta)/k$.
Note the rise of the curves as $k \rightarrow \Lambda=8$. This is consistent
with the effective range
$r_0=0.25$ being fairly close to the unitary limit $r_0=0$ whence the
potential is singular at $k=\Lambda$ as shown by eq. (\ref{vpi2}). (See
also Fig. \ref{wavnewpot}).

\section{Summary and Conclusions}
Our previously reported  ladder calculation of the total energy 
of a unitary gas \cite{hsk08} gave the
result $\xi=0.24$ (energy in units of kinetic energy of the uncorrelated
gas). The unitary interaction  (\ref{vpi2})  used in that
calculation was obtained by 
inverse scattering and it satisfys the unitary condition 
exactly by having scattering
phase-shifts $\delta(k)=\pi/2$ for all $k<\Lambda$, the momentum cut-off.
Although  not claimed to be an 'exact' many-body calculation it had the
virtue of being an almost completely analytic result. It also relied to
some extent on the succesful Brueckner approach to the nuclear many-body
problem, being a ladder-summation and respecting the Fermi-statistics of
the problem. 
It did however neglect another aspect of the Brueckner method,
the mean-field (self-energy) insertions in propagator-lines. 

But an important result of the previous calculations was also that  
the energy  increases
rapidly with increasing value of $r_0$,  while keeping the scattering length
large.  
This is also   shown here by the lowest curve  in Fig. \ref{unir0} above.

In the present report two separate issues were adressed. One is the
non-uniqueness of interactions derived from the on-shell scattering
phaseshifts. The other is the previously neglected insertions in
hole-lines. 

Regarding the first issue it is basically unknown territory. 
The 'potential' is not an observable, but a theoretical construction.
This is for example illustrated by the nucleon-nucleon  $^1S_0$-potentials
shown in Fig. (3) of ref. \cite{hsk05} all fitting low-energy phase-shifts
but  differ with  $\Lambda$. 
In the present calculations the interactions are assumed to be purely
2-body. In the real world, e.g some atomic gas, it may be necessary to
introduce higher order forces 
to allow for 'distortions' of the force-field due to neighboring
particles. The results shown in Fig. \ref{unioff} 
illustrate the fact that phase-shifts alone do not define the properties of
a many-body system. Fit to phase-shifts may be considered a necessary
condition in this type of many-body theory.
But it
is not a sufficient condition to predict the properties of a many-body system.
The caculations presented here fulfill the first, the necessary 
condition but considerable variations in energy  can be seen in Fig.
\ref{unioff} if off-shell properties are changed. The unsolved problem is
of course how to relate these to specific gases. 
We may however also argue that in the unitary limit 
the rank-1 potential (\ref{vpi2}) fulfills 
both conditions, the correct on- as well as off-shell dependence.
As referred to above, it has already been demonstrated that the nuclear
$^1S_0$ interaction with considerable success can be represented 
by a separable potential. This could be attributed to ae nearly bound
state in this case.
In the unitary limit the pole of the
scattering matrix lies not only close to but exactly on the real axis. 
The interaction should therefore in this case be an even better 
candidate for being represented by a separable potential at least 
in the vicinty of this pole. (See e.g.  ref. \cite{gebrown} regarding this
subject).
Question is, is the separable interaction given by eq. (\ref{vpi2}) 
unique. It certainly is if one restricts to small momenta.

As to the the mean-field propagation in the definition of the 
effective in-medium interaction,
it is of utmost importance in the nuclear
many-body system and it is included in the definition of the Brueckner reaction
matrix, eq. (\ref{BK}). It is the main reason for nuclear saturation in
Brueckner theory of nuclei, and that is
mainly a consequence of the  momentum-dependent mean field $U(k)$ so that  
$\Delta U \neq 0$.  The  result of this is that the dispersion correction, 
eq. (\ref{disp}) is large and increasing with density.

The situation is different for the case studied here. The mean field $U(k)$ is
nearly constant for the unitary short-ranged interaction. 
The mean field is practically independent of momentum so that 
$\Delta U$ in eq. (\ref{disp}) would be  nearly zero with a \it continuous
\rm choice. There are however other
facts to consider here. The first is that $I_w$ in eq. (\ref{disp}) 
increases as $r_0$ decreases  approaching the unitary limit. 
Referring to works by Brueckner, Gammmel, Bethe and others it was pointed
out in Sect. 2  that the determination  of  $\Delta U$ involves an effort to sum
higher order terms in the calculation of the energy. 
The mean-field $U_h(k)$ for $k<k_f$, to be used for
insertions in hole-lines should be calculated from $K$ as defined by eq.
(\ref{BK}).   Regarding $U_p(k)$ with $k>k_f$ the situation is not that
clear.  Bethe \cite{bet65} and others did for example suggest to choose
$U_p(k)=0$,  in the literature often referred to as the \it standard \rm choice.

The effect of hole- and particle- insertions was investigated above. 
Referring to Fig. \ref{unir0} one notes that without any insertions 
the enrgy  decreases rapidly as $r_0$ is decraesed agreeing with an earlier
report.\cite{hsk08} One finds on the other hand a dramatic opposite effect
with the standard choice. 
By reducing the mean field by a factor of $\frac{1}{4}$ one sees that
$\xi=0.4 \Leftrightarrow 0.5$ and almost independent of $r_0$.

These results do not conclusively suggest a definite value for $\xi$.
They only suggest that the previously reported value of $\xi$
should be modified due to the mean field and perhaps off-shell effects.

The final conclusion is 
that there may not after all exist any definite disagreement 
between these type of calculations (inspired by the success of Brueckners 
many-body theory) and the (in principle) more accurate Monte-Carlo 
and other methods.
The present investigation rather indicates that the basic
ideas of Brueckner theory, successful as a theory of nuclei  may also be
carried over to the unitary system although higher order terms, not
customarily included in "Brueckner Theory" would have to be included. 
The relatively strong correlations expressed by the large effects of mean field
insertions and consequently large values of $I_w$ suggest that such higher order
terms may be important.

Brueckner type calculations with results similar to the Monte Carlo
were also reported by Siu et al\cite{siu08,don10}.
They differ from the present
by including also ring diagrams and by the use of different interactions.
Although there is some agreement there is at this time
not enough ground for a direct comparison of results.

\end{document}